\documentclass[journal=nanoletters,manuscript=letter]{achemso}

\usepackage{chemformula} 
\usepackage[T1]{fontenc} 
\usepackage{amsmath}
\usepackage{amssymb}
\DeclareUnicodeCharacter{2009}{\,} 

\author{Arjav Shah}
\affiliation{Department of Chemical Engineering, Massachusetts Institute of Technology, Cambridge, MA 02142, United States}
\alsoaffiliation{Singapore-MIT Alliance for Research and Technology Centre, Singapore 138602, Singapore}
\author{Shakul Pathak}
\affiliation{Department of Chemical Engineering, Massachusetts Institute of Technology, Cambridge, MA 02142, United States}
\author{Slaven Garaj}
\affiliation{Department of Physics, National University of Singapore, Singapore 119077, Singapore}
\alsoaffiliation{Singapore-MIT Alliance for Research and Technology Centre, Singapore 138602, Singapore}
\author{Martin Z. Bazant}
\affiliation{Department of Chemical Engineering, Massachusetts Institute of Technology, Cambridge, MA 02142, United States}
\alsoaffiliation{Department of Mathematics, Massachusetts Institute of Technology, Cambridge, MA 02142, United States}
\author{Ankur Gupta}
\affiliation{Department of Chemical and Biological Engineering, University of Colorado, Boulder, CO 80303, United States}
\author{Patrick S. Doyle}
\email{pdoyle@mit.edu}
\affiliation{Department of Chemical Engineering, Massachusetts Institute of Technology, Cambridge, MA 02142, United States}
\alsoaffiliation{Singapore-MIT Alliance for Research and Technology Centre, Singapore 138602, Singapore}

\title{A universal approximation for conductance blockade in thin nanopore membranes}

\abbreviations{IR,NMR,UV}

\begin{document}








\begin{abstract}
Nanopore-based sensing platforms have transformed single-molecule detection and analysis. The foundation of nanopore translocation experiments lies in conductance measurements, yet existing models, which are largely phenomenological, are inaccurate in critical experimental conditions such as thin and tightly fitting pores. Of the two components of the conductance blockade, channel and access resistance, the access resistance is poorly modeled. We present a comprehensive investigation into the access resistance and associated conductance blockade in thin nanopore membranes. By combining a first-principles approach, multi-scale modeling, and experimental validation, we propose a unified theoretical modeling framework. The analytical model derived as a result surpasses current approaches across a broad parameter range. Beyond advancing theoretical understanding, our framework's versatility enables analyte size inference and predictive insights into conductance blockade behavior. Our results will facilitate the design and optimization of nanopore devices for diverse applications, including nanopore base calling and data storage.
\end{abstract}

Nanopores have evolved as valuable single-molecule analytical tools over the past four decades. While ubiquitous in nucleic acid sequencing applications, the utilization of nanopores extends into broader molecule sensing, biomolecule characterization, catalysis, desalination, and power generation.\cite{ying_nanopore-based_2022,muthukumar_single-molecule_2015, wang_nanopore_2021, heiranian_water_2015, macha_2d_2019, henrique_charging_2022}  The technology has been widely exploited for the detection, fingerprinting, and sequencing of biomolecules for advances in medicine, biotechnology, and forensics. In particular, it has been extensively used to probe the mechanics and dynamics of DNA including dsDNA\cite{deamer_three_2016} and DNA knots \cite{plesa_direct_2016, sharma_complex_2019,sharma_dna_2021}, proteomics \cite{yusko_real-time_2017, brinkerhoff_multiple_2021, alfaro_emerging_2021}, virology\cite{Uram2006, mcmullen_stiff_2014, li_metrology_2023, taniguchi_combining_2021}, etc..

\begin{figure}[!ht]%
\centering
\includegraphics[width=1.0\textwidth]{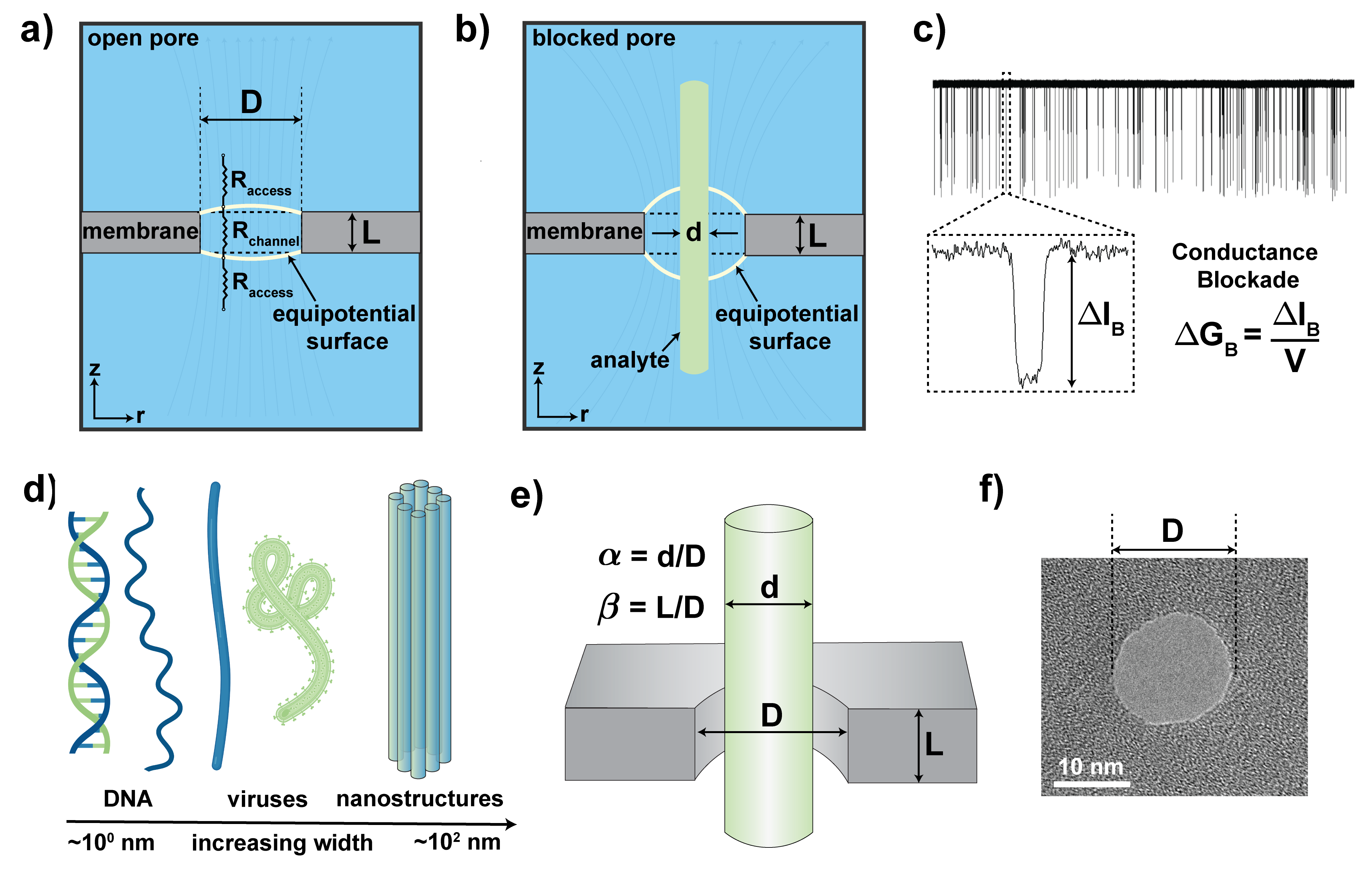}
\caption{State of the nanopore system corresponding to the changes in the curvature of equipotential surface for \textbf{a)} open pore shown with minimal circuit representation of the system, and \textbf{b)} blocked pore in the presence of a long, cylindrical analytes. \textbf{c)} Representative resistive pulse outputs from experiments with amplitude current blockade $\Delta I_B$. \textbf{d)} Analytes of interest ranging from DNA, viruses, and nanostructures. \textbf{e)} Dimensionless groups in the system, $\alpha$ (analyte-to-pore diameter ratio) $\in [0,1)$ and $\beta$ (pore aspect ratio) $\in (0,5)$ typically. \textbf{f)} TEM image of a silicon nitride  SiN$_{\text{x}}$ pore used for detection.} \label{fgr:1}
\end{figure}

Mapping the experimental observables to their associated analyte characteristics is a critical enabler for nanopore sensing in diverse applications. Conductance measurement is the cornerstone of nanopore translocation experiments. A baseline \textit{open} current exists due to the electrophoretic movement of ions through a pore in an otherwise insulating free-standing membrane (Figure \ref{fgr:1}(a)). The presence of an analyte inside the pore impedes the flow of ions. As a result, the translocation of a charged analyte through the pore under the influence of an electric field causes a transient change in the electrical resistance of the system (Figure \ref{fgr:1}(b)). Based on the principle of \textit{resistive pulse sensing}, the change in resistance manifests as a modulation in the ionic current in current-time pulse signatures in experiments (Figure \ref{fgr:1}(c)) \cite{kasianowicz_characterization_1996}. The nanopore system is ultrasensitive to these changes in the ionic current such that the electrical signatures encompass information about the morphological characteristics of translocating analytes. For instance, the amplitude of a signature pulse, known as \textit{current blockade} $\Delta I_B$, encodes analyte width information, which is important for molecule characterization and fingerprinting applications. Normalized $\Delta I_B$ is referred to as \textit{conductance blockade} $\Delta G_B$ (Figure \ref{fgr:1}(c)). An accurate, physics-based model of $\Delta G_B$ is critical for the interpretation of experimental observables. 

Long, rod-like, charged cylindrical objects such as dsDNA, filamentous viruses such as \textit{fd} or TMV, nanorods, and DNA nanostructures such as DNA origami bundles, etc. (Figure \ref{fgr:1}(d)) have been studied extensively because of their importance in applications such as  sequencing, diagnostics, and biophysical studies. \cite{mcmullen_stiff_2014, li_metrology_2023, Wu2016, qiao_capture_2020, venta_gold_2014, he_liqun_dna_2023} In addition to the prevalence of such analytes, their cylindrical symmetry offers mathematical convenience for the use of cylindrical or oblate spheroidal coordinate systems in analytical conductance models. Relevant length scales for a system with cylindrical analytes including the pore diameter $D$, membrane thickness $L$, and analyte diameter $d$ are depicted in Figure \ref{fgr:1}(e,f). The analyte length $l$ does not play a role in modulating $\Delta G_B$ as long as $l \gg L$. Based on these three length scales, two dimensionless groups are defined in the system: $\alpha = d/D$ represents how loosely/tightly fit the analyte is inside the pore, and $\beta = L/D$ is the pore aspect ratio (Figure \ref{fgr:1}(e)). 

The total conductance of the nanopore system is determined by the sum of key resistances in series which include the access (or convergence) and channel (or pore) resistances. The resistance to the flow of ions inside the pore channel is referred to as the channel resistance, $R_{channel}$. The potential gradient across the bulk electrolyte solution due to the distortion and convergence of electric field lines around the pore mouths is approximated by access resistance, $R_{access}$; equivalent to entrance effects in fluid flow through constrictions. Existing conductance models have been primarily developed for long, cylindrical analytes \cite{kowalczyk_modeling_2011, carlsen_interpreting_2014, tang_understanding_2022}. While these models work well for small $\alpha$ and $\beta \gg 1$, they fail for thin membranes ($\beta \ll 1$). However, with the advent of and progress in two-dimensional materials, the increasing use of extremely thin membranes for nanopore experiments is imperative. These are ideal for high-resolution, high-throughput nanopore-based single-molecule detection. They have been reported to have several advantages in terms of manufacturability, better signal-to-noise ratio (SNR), electrical sensitivity, and stability.\cite{arjmandi-tash_single_2016, danda_two-dimensional_2019} In addition, tightly-fitted pores ($\alpha \rightarrow 1$) are critical for high experimental sensitivity and SNR.\cite{danda_two-dimensional_2019} As a result, accurate models for conductance in thin and tightly-fitted pores, \textit{i.e.}, $\Delta G_B$ for $\alpha \rightarrow 1$ and $\beta \ll 1$, are critical. 

\begin{figure}[!ht]%
\centering
\includegraphics[width=1.0\textwidth]{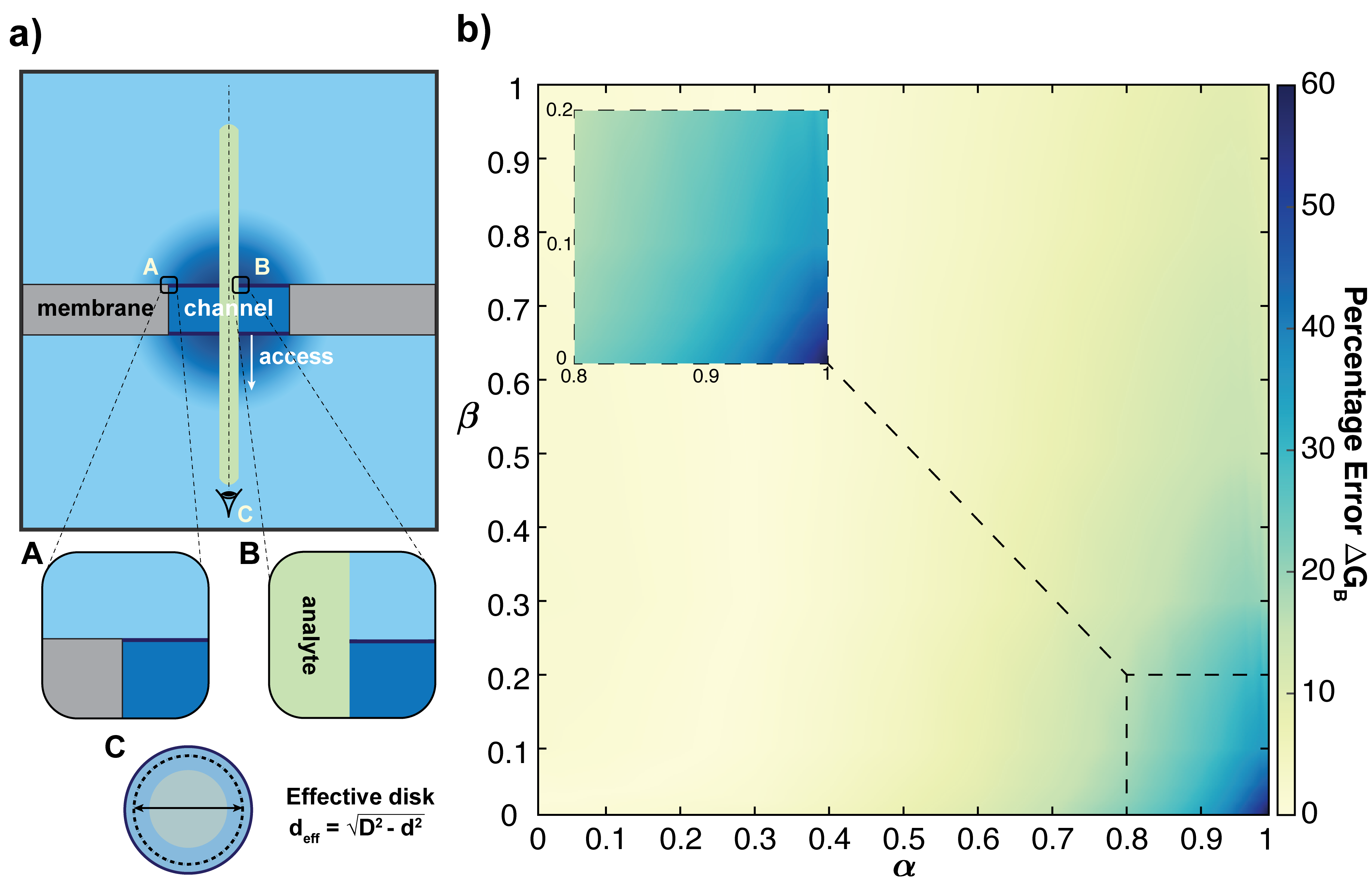}
\caption{\textbf{Kowalczyk Model -} \textbf{a)} System partitioning into channel and access regions based on a flat equipotential surface (highlighted in \textit{A} and \textit{B}) which is treated as an effective disk shown in \textit{C}. \textbf{b)} Heatmap of relative error in conductance blockade $\Delta G_B$ as compared to fully-coupled, continuum simulations. \textit{Inset}: highlighting the parameter space with significant errors in $\Delta G_B$, as high as 60\%.}\label{fgr:2}
\end{figure}

A key component of $\Delta G_B$, $R_{access}$ dominates for thin pores ($R_{access}/R_{channel}\sim 1/\beta$; see SI). Based on past works, it is widely understood that $R_{access}$ for an \textit{open pore} depends only on the resistivity of the conducting medium and pore diameter $D$\cite{maxwell_treatise_1881, hille_pharmacological_1968, hall_access_1975}. However,  a comprehensive understanding of the perturbation in $R_{access}$ in the presence of an analyte inside the pore remains elusive. First-generation models for $R_{access}$, proposed by Hall and Hille, have been developed for ion channels in pharmacological applications, primarily dealing with open pore scenarios ($\alpha = 0$).\cite{hille_pharmacological_1968, hall_access_1975} The presence of an analyte has only been {phenomenologically} or empirically accounted for thus far, including in more recent models by Kowalczyk \cite{kowalczyk_modeling_2011} and Carlsen\cite{carlsen_interpreting_2014}. The most widely-used model is that of Kowalczyk which combines $R_{access}$ from Hall\cite{hall_access_1975} and $R_{channel}$ from Ohm's Law integrated across the channel. The presence of an analyte is captured based on an effective pore diameter which is exact for $R_{channel}$ but is only a first-order approximation for $R_{access}$ (Figure \ref{fgr:2}(a)). The Carlsen model overcorrects the Kowalczyk model by accounting for the analyte as an added negative access conductance in addition to the effective pore diameter.\cite{carlsen_interpreting_2014}

The most important assumption in Hall's approach is that the equipotential surface at the pore mouth is a planar disk (Figure \ref{fgr:2}(a), section A,B). While this assumption is suitable for an open pore, it breaks down for tightly-fitted pores. For an open pore, a large fraction of electric field lines can pass straight through the pore unhindered. In contrast, in the presence of an analyte, the field lines are no longer straight in the pore channel and bend significantly around the pore mouth as they exit. As a result, the equipotential surface at the pore mouth cannot be a flat disk as assumed in the model since it has to remain perpendicular to the field lines. Moreover, in the presence of an analyte, the pore is no longer `open' and thus not a complete disk anymore (Figure \ref{fgr:2}(a), section C).

To overcome the assumptions of the existing models, we use fully-coupled simulations of the Poisson-Nernst-Planck and Navier-Stokes equations with appropriate boundary conditions in \textit{COMSOL Multiphysics}. We show that the state-of-the-art approaches perform poorly in the critical experimental limits ($\alpha\rightarrow 1, \beta\ll 1$). The heatmap in Figure \ref{fgr:2}(b) depicts the relative error in the prediction of $\Delta G_B$ using the Kowalczyk model when compared to the simulation results. The Kowalczyk model has errors as high as 60\% for thin and tightly-fitted pores ($\alpha\rightarrow 1, \beta\ll 1$). Unsurprisingly, these are the cases farthest from open pore scenarios where the flat disk assumption is not justified. This limits the usage of the existing models in current applications. \cite{tang_understanding_2022}

We hypothesize that an annular disk is a better approximation than an effective disk, and that the equipotential surface is an elliptical arc, drawing inspiration from analogous heat and mass transport problems.\cite{wexler_bending_2013, atwal_mass_2021} The access region and hence, its associated resistance, $R_{access}$ is perturbed \textit{non-trivially} during the translocation of an analyte through the region. To test our hypotheses, we use a combination of simulations, analytical modeling, and experimental validation to elucidate the role of access region in determining $\Delta G_B$ in the presence of cylindrical analytes, particularly for thin pores.

\begin{figure}[!ht]%
\centering
\includegraphics[width=1.0\textwidth]{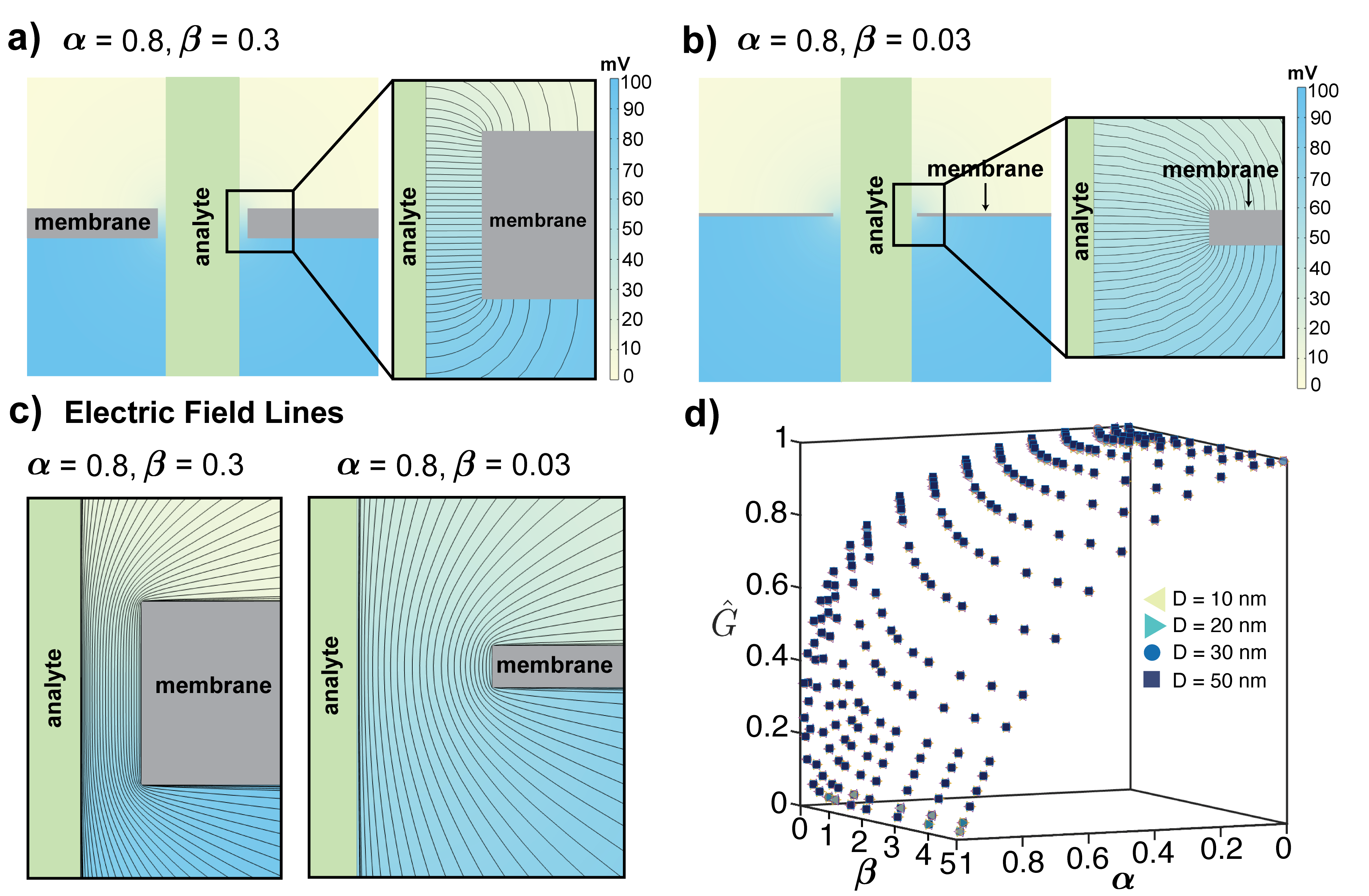}
\caption{\textbf{Continuum Simulations - } \textbf{a)} and \textbf{b)} Electric potential gradient and equipotential surfaces (enlarged) for thick and thin pores, respectively. \textbf{c)} Bending of electric field lines around the pore mouth. \textbf{d)} Variation of dimensionless conductance blockade (equation S9) over $\alpha$-$\beta$ space shown for various $D$ values.}\label{fgr:3}
\end{figure}

The equipotential surface at the pore mouth is curved (Figure \ref{fgr:3}(a,b)) since the fraction of electric field lines that bend for thin membranes and tightly-fitted pores is very high (Figure \ref{fgr:3}(c)). As a result, the assumption of a planar disk at the pore mouth is a poor approximation in this limit and influences the accuracy of the existing $R_{access}$ models. In addition, due to the assumption of a hyperbolic shape of analyte (specifically dsDNA) in the pore instead of a cylinder, the Kowalczyk model consistently overestimates $\Delta G_B$.  

Towards proposing a new modeling framework and the development of a more accurate analytical conductance model, we show that two dimensionless groups, $\alpha$ and $\beta$, are sufficient to predict the conductance of a nanopore system, \textit{i.e.}, if two systems have the same ($\alpha$, $\beta$), they will have the same non-dimensional conductance, $\hat G = G_{B,total}/G_{O,channel}$ (equation S9), where $G_{O,channel}$ and $G_{B,total}$ refer to the open channel and total blocked pore conductances, respectively. This is evident by a collapse of data points with the same ($\alpha, \beta$) values over the entire parameter space in Figure \ref{fgr:3}(d). Leveraging this insight,  we propose and validate a universal, physics-inspired analytical model for $\Delta G_B$, demonstrating superior performance compared to the existing models for a wide parameter range. We present a modeling framework with three key novel features: (1)
accurate estimation for capacitance of an annular disc and hence, $R_{access}$ dominant for thin pores, (2) non-hyperbolic, cylindrical analyte shapes, and (3) incorporation of electric field bending and hence, the dependence of channel and access region partitions on both $\alpha$ and $\beta$.

\begin{figure}[!ht]%
\centering
\includegraphics[width=1.0\textwidth]{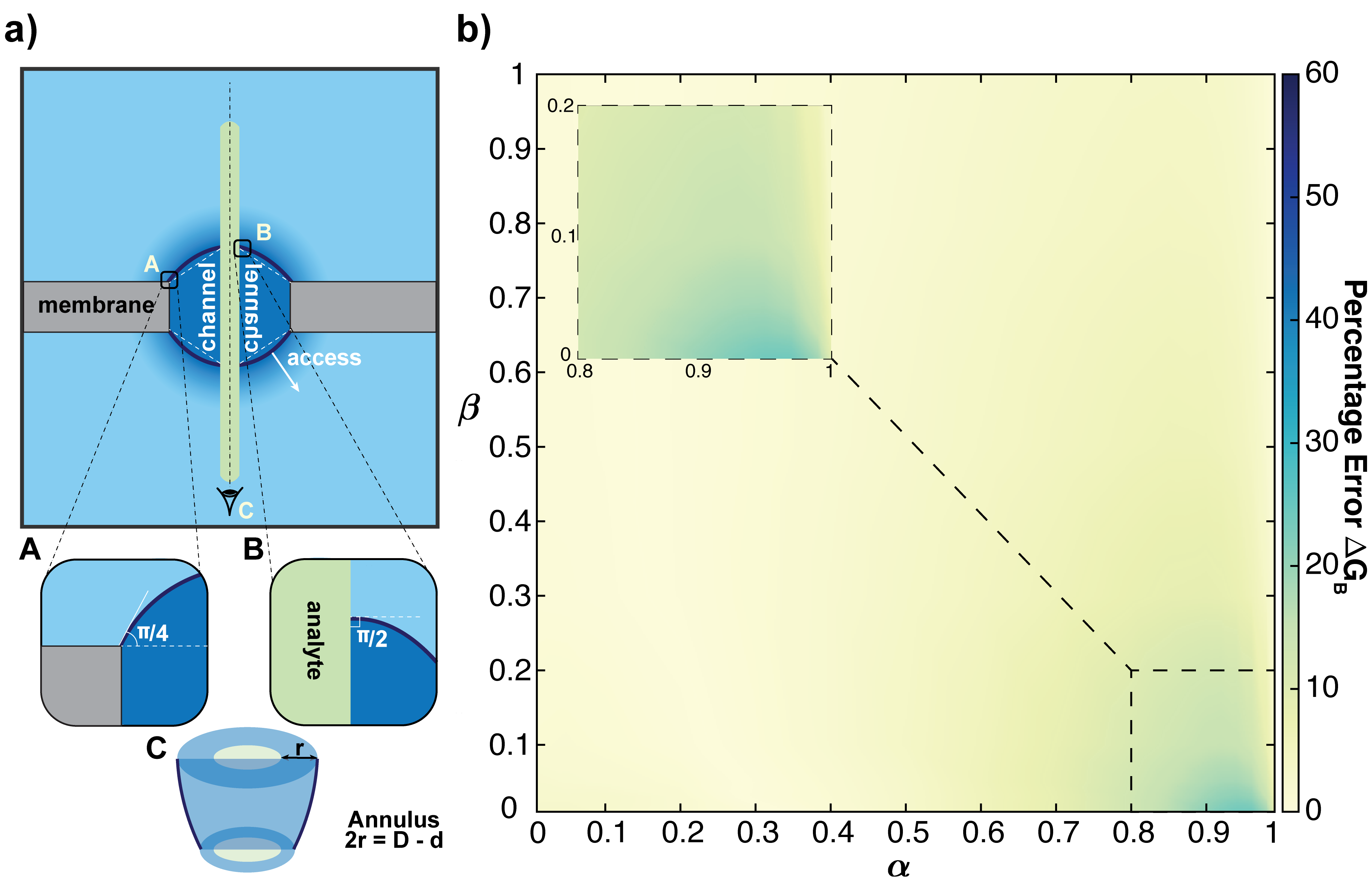}
\caption{\textbf{Our Model -} \textbf{a)} Updated system partitioning into channel and access regions based on equipotential surface (highlighted in \textit{A} and \textit{B}) which forms a truncated elliptical cone with an annular region shown in \textit{C}. \textbf{b)} Heatmap of relative error in conductance blockade $\Delta G_B$ as compared to fully-coupled, continuum simulations. \textit{Inset}: emphasizing the significant reduction in errors in $\Delta G_B$ for an important parameter range.}\label{fgr:4}
\end{figure}

Consider a long, cylindrical analyte of diameter $d$ with length $l$ long enough to pierce both the access regions of a nanopore of diameter $D$ and thickness $L$. We focus on concentrated electrolyte systems and any double-layer effects can be neglected to first order (Debye length, $\lambda_{DL}/D \ll 1$ and $\lambda_{DL}/d \ll 1$; for \textit{e.g.}, 0.3 nm for 1 M KCl). Therefore, any surface charge on the analyte or pore is screened by a thin double layer. At the pore length scale, the analyte and pore surfaces can be assumed to act like an insulator in a background electric field. Charge screening in the pore results in field lines running parallel to the pore wall and virus surface (Figure \ref{fgr:3}(c)). The geometric boundaries between the channel and access region can now be assigned by noting that the equipotential line that passes through the corner of the pore wall (Figure \ref{fgr:4}(a), section A) is incident on the corner at an angle of 45$^\circ$ (see SI).\cite{olver_complex_2017} Since the analyte surface has no net charge at high ionic strengths, the equipotential lines are incident perpendicular to its surface (Figure \ref{fgr:4}(a), section B). 

Similar to Hall's approach, we seek to find the capacitance of the equipotential surface sketched in Figure 4(a). \cite{hall_access_1975} As a first-order approximation, we consider the equipotential surface as a flat punctured disk surface of inner diameter $d$ and outer diameter $D$ although it is a surface of revolution of an elliptic arc \cite{atwal_mass_2021}. To estimate the capacitance, we utilize the work of Smythe who proposed an iterative method for this problem, starting with the charge distribution of an equipotential disk (see SI) \cite{smythe_capacitance_1951}. These calculations yield:
\begin{equation}\label{eqn:cannulus}
    \frac{C_{annulus}}{C_{disk}} = \frac{2}{\pi}\left[\cos ^{-1}\alpha+\left(1-\alpha^2  \right)^{\frac{1}{2}} \tanh ^{-1}\alpha\right]\left(1+\left(\frac{0.0143}{\alpha}\right) \tan ^3\left(1.28{\alpha}\right)\right)
\end{equation}
where $C_{annulus}$ and $C_{disk}$ are the capacitances of flat annulus and disk, respectively. For $\alpha \rightarrow 0$, we recover that the values are equal. Hereafter, we refer to the open and blocked pore conductances as $G_{O,i}$ and $G_{B,i}$, respectively, such that $i \in \{total, access, channel\}$. To obtain the desired quantity $G_{B,access}$, we invoke that the conductance is proportional to the capacitance ($G = C\sigma/\varepsilon$, where $\sigma$ and $\varepsilon$ are the electrolyte conductivity and permittivity of free space, respectively) and hence, 
\begin{equation}\label{eqn:Gannulus}
    \frac{G_{B,access}}{G_{O,access}} = \frac{2}{\pi}\left[\cos ^{-1}\alpha+\left(1-\alpha^2 
    \right)^{\frac{1}{2}} \tanh ^{-1}\alpha\right]\left(1+\left({0.0143}{\alpha}\right) \tan ^3\left(1.28\alpha\right)\right)
\end{equation}

The other contribution to the total conductance, $G_{B, channel}$, is identical to the previous approaches for the flat punctured disk surface: 
\begin{equation}\label{eqn:Gch}
    G_{B,channel} = \frac{\pi\sigma D}{4\beta}\left(1 - \alpha^2\right)
\end{equation}
The conductance blockade $\Delta G_B$ can be obtained by subtracting the total blocked conductance from the open conductance:
\begin{equation}\label{eqn:Gtot}
  \Delta G_{B} = G_{O,total} - G_{B,total} = 
    \sigma D \left(\frac{4\beta}{\pi}+1\right)^{-1}- (G_{B,channel}^{-1} + G_{B,access}^{-1})^{-1} 
\end{equation}


As a second-order approximation, the equipotential surface is now assumed to be an inclined annular disk since the equipotential surface that passes through the corner of the pore is curved (Figure \ref{fgr:3}(b,c)). Assuming a partitioning scheme where the boundary between the access and channel region is given by a straight line inclined at an angle $\theta$ with the horizontal, instead of fixing it at 45$^\circ$ as previously discussed (Figure \ref{fgr:4}(a)), yields an improved estimate for an appropriately fitted value of the parameter $\theta$ (second-order correction); such that $\theta$ is the only phenomenological parameter. The updated model for $G_{B,access}$ reads (see SI):
\begin{align}\label{eqn:Gaclb}
    G_{B,access} = \frac{\sec \theta \sigma}{\varepsilon}C_{annulus} 
\end{align}
Proceeding similar to equation \ref{eqn:Gch}, we obtain equation \ref{eqn:Gchlb} (see SI). It is noteworthy that the dimensional prefactors in both these equations scale as $\sigma D/\beta$. 
\begin{align}\label{eqn:Gchlb}
    G_{B,channel} = \frac{\pi \sigma D}{2\beta}\left(-(1-\alpha) + \left(1+\frac{\beta}{tan\theta}\right)\ln\left(1+\frac{tan\theta }{\beta}(1-\alpha)\right)\right) 
\end{align}

We systematically evaluate the model performance (equations \ref{eqn:Gtot}-\ref{eqn:Gchlb}) against simulation results for a wide range of parameters: $\alpha \in [0,0.99]$ and $\beta \in [0.03, 5]$ for $D \in [10,50]$ nm, equating to 867 unique cases. Figure \ref{fgr:4}(b) shows the relative error in $\Delta G_B$ estimated using our newly proposed model, which upon comparison with Figure \ref{fgr:2}(b), clearly underscores that our new model significantly outperforms the state-of-the-art model in the thin and tightly-fitted pore limit, i.e., small $\beta$ and large $\alpha$. Combining all cases with $\beta \leq 0.1$, average error: 9\% and 18\%; maximum error: 24\% and 61\% for our and Kowalczyk models, respectively (table S2). Additional analyses and comparisons for alternative models are shown in Figures S2 and S3 \cite{kowalczyk_modeling_2011, carlsen_interpreting_2014}. 

\begin{figure}[!tbh]%
\centering
\includegraphics[width=0.95\textwidth]{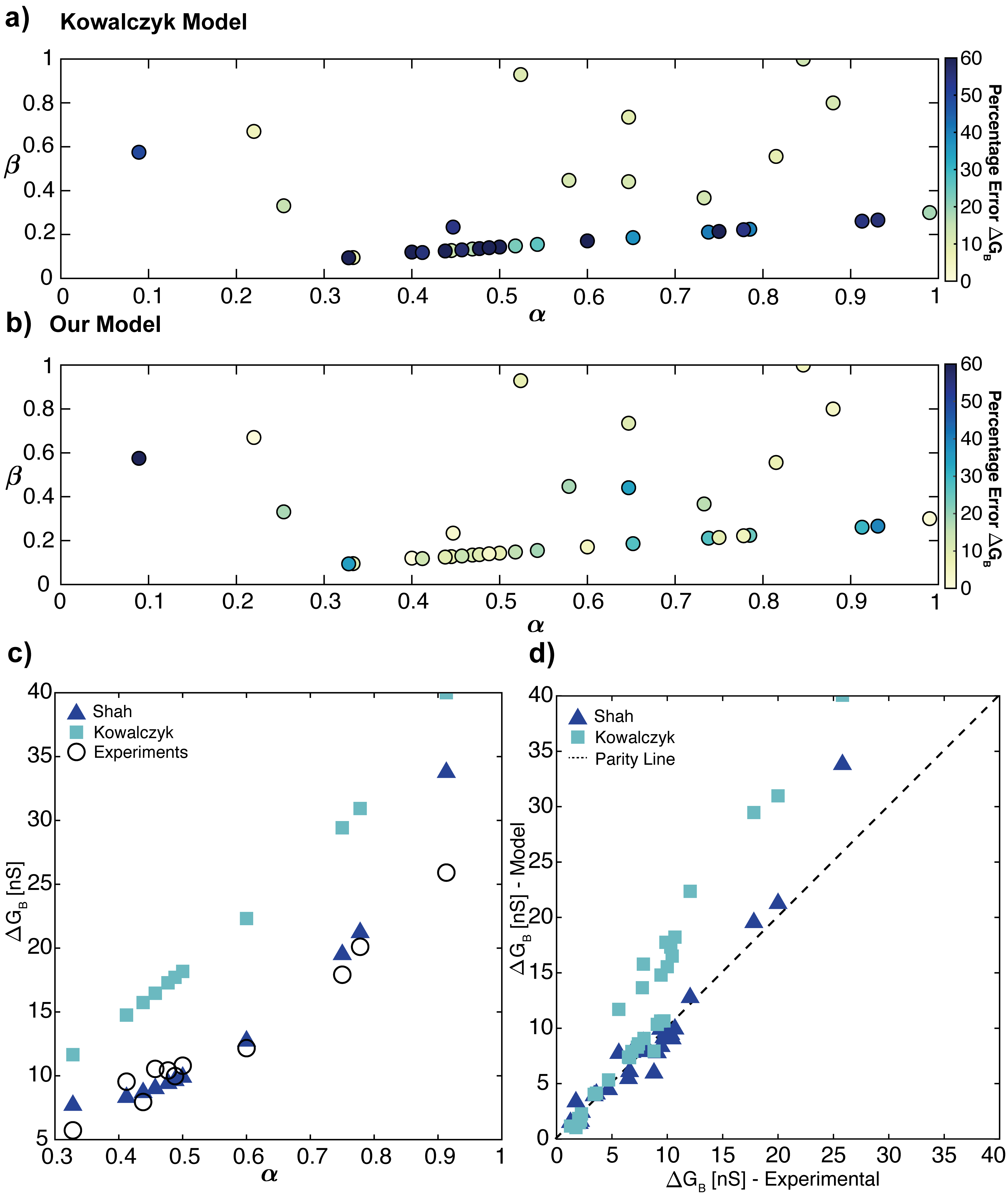}
\caption{\textbf{Experimental Validation -} Comparison of relative errors in conductance blockade $\Delta G_B$ estimates from \textbf{a)} Kowalczyk model and \textbf{b)} our model (Shah) with respect to experiments \cite{ garaj_graphene_2010, garaj_molecule-hugging_2013, liu_boron_2013, wang_fundamental_2017, merchant_dna_2010, rodriguez-manzo_dna_2015, schneider_dna_2010}.  \textbf{c)} Comparison of absolute $\Delta G_B$ values for $\beta \sim 0.2$ for our model, Kowalczyk's model and experiments. \textbf{d)} Parity plot comparing the model results with experimental data, highlighting the superior performance of the our proposed model over the state-of-the-art conductance model.}\label{fgr:5}
\end{figure}


Our new model, in addition to fully-coupled simulations, successfully predicts experimental data over a wide range of parameters reported by different research groups. A comprehensive summary of the data can be found in the SI (table S1). \cite{ garaj_graphene_2010, garaj_molecule-hugging_2013, liu_boron_2013, wang_fundamental_2017, merchant_dna_2010, rodriguez-manzo_dna_2015, schneider_dna_2010} An extensive experimental validation for the new model is shown in Figure \ref{fgr:5}. Figures \ref{fgr:5}(a) and \ref{fgr:5}(b) present the relative error in $\Delta G_B$ prediction from Kowalczyk's and our new model when compared to the reported experimental data. The average percentage errors over 44 data points are 36\% and 16\% for the two models respectively. It is clear from these figures that the new physics-based approach yields a model that outperforms the existing state-of-the-art models over the complete range of $\alpha$ for thin pores. Moreover, upon comparing the absolute $\Delta G_B$ values in Figure \ref{fgr:5}(c) for a particular dataset that reports a fairly constant $\beta$, the markers for the new model (solid triangles) are much closer to the experimental markers (open circles) as $\alpha$ is varied, suggesting a better match. Finally, the validity and higher accuracy of the new model become abundantly clear as the predictions lie close to the parity line $(x=y)$ in Figure \ref{fgr:5}(d). 

The results are not sensitive to the phenomenological parameter $\theta$ in our physics-based model (Figure S6(b)) which is evaluated as a best-fit parameter using a fourth of the simulation dataset. We show that even without an added second-order approximation ($\theta = 0$, i.e., flat equipotential surface) the model (equations \ref{eqn:Gannulus}-\ref{eqn:Gtot}) outperforms the existing models (Figure S6(a)). The curvature of the equipotential surface is already accounted for, in part, by equation \ref{eqn:Gannulus}. 

In conclusion, the new model serves the need for a more accurate model in the especially critical thin pore limit. It demonstrates the non-trivial dependence of $R_{access}$ on the presence of an analyte and successfully predicts conductance for a wide range of the experimental dataset. Going beyond current applications, the model formulation can be adopted for problems involving any gradient-driven flow such as diffusiophoresis or pressure-driven flows. The general framework for solving the Laplace equation lends the approach its broader applicability. Fundamentally, our approach building on Smythe's work is powerful since it can be extended beyond axisymmetric geometries to systems that admit Schwartz-Cristoffel transformation.\cite{driscoll_schwarz-christoffel_2002, bazant_conformal_2004} 

We propose a novel partitioning scheme for the nanopore system accounting for the changes in the equipotential surface as a result of analyte translocation through nanopores. A physics-based, more accurate conductance model emerges as a result of the improved system decomposition and capacitance calculations. The model has been developed for high salt conditions, where the diffusive fluxes are negligible in strength as compared to the deterministic migration fluxes caused by the applied electric field. The majority of nanopore measurements are conducted at high salt conditions to enhance detection sensitivity and signal-to-noise ratio, precisely the conditions that the new model is expected to perform the best at. Overall, we expect our modeling framework to become a convenient yet accurate method for conductance estimates for various pore geometries and driving forces in nanopore translocation.

We anticipate that these results can be directly applied in various solid-state nanopore applications including data storage \cite{chen_nanopore-based_2020}, profiling polymer topology \cite{sharma_complex_2019, sharma_dna_2021, rheaume_nanopore_2023},  peptide sequencing \cite{alfaro_emerging_2021}, and virus screening among others. The modeling framework can also be used to improve the accuracy of nanopore base calling in genomic technologies. The state-of-the-art sequencing technologies suffer from high error rates of 5-15\% \cite{rang_squiggle_2018}. Such a framework can support model development to overcome these challenges by helping us better understand the signal characteristics as well as for signal calibration and normalization. Accurate conductance models like ours can improve the accuracy and reliability of various single-molecule applications of nanopores\cite{wang_nanopore_2021}.
\begin{suppinfo}
Supporting Information: Additional model results and comparison, experimental validation data, methods including analytical model formulation, simulation model setup. The following file will be available upon request from the authors. 
\begin{itemize}
   \item Supplementary Information (SI\_Shah\_ConductanceModel.pdf)
 \end{itemize}
\end{suppinfo}

\begin{acknowledgement}
The authors thank Rohit Karnik for fruitful discussions and Kun Li for the nanopore experimental pulses as well as pore TEM image. This work is supported by the National Research Foundation, Prime Minister's Office, Singapore under its Campus for Research Excellence and Technological Enterprise (CREATE) program, through the Singapore MIT Alliance for Research and Technology (SMART): Critical Analytics for Manufacturing Personalised-Medicine (CAMP) Inter-Disciplinary Research Group. A.S. acknowledges the support from the MathWorks Engineering Fellowship. A.S. and P.S.D. acknowledge the MIT SuperCloud and the Lincoln Laboratory Supercomputing Center for providing HPC resources that have contributed to the simulation results reported within this paper. A.G. thanks the National Science Foundation (CBET - 2238412) CAREER award for financial support.
\end{acknowledgement}

\section{Author Information}
\subsection{Corresponding Author}
\textbf{Patrick S. Doyle} -
Department of Chemical Engineering, Massachusetts Institute of Technology, Cambridge, MA 02142, United States; Singapore-MIT Alliance for Research and Technology Centre, Singapore 138602, Singapore; Email: pdoyle@mit.edu

\subsection{Author Contributions}
A.S., S.G., and P.S.D. conceived the project. A.S. performed the simulations. A.S., S.P., A.G., and P.S.D. analyzed the data. A.S., S.P., M.Z.B., A.G., and P.S.D. developed the analytical model. A.S., S.P., and S.G. performed the experimental validation. All authors discussed the results and wrote the manuscript. All authors have given approval to the final version of the manuscript.

\subsection{Notes}
The authors declare no competing financial interest.

\bibliography{ms} 

\end{document}